# Dynamics of the temperature regime of permafrost soil in the vicinity of the main gas pipeline taking into account climate warming

A.A. Fedotov, P.V. Khrapov, A.E. Dengovskaya

*Abstract* – **An initial-boundary value problem for an unsteady two-dimensional heat conduction equation in a bounded domain modeling the unsteady temperature distribution of permafrost soil in the vicinity of a main gas pipeline, taking into account climate warming, is investigated. The parameters of the mathematical model are selected in accordance with experimental data on gas transportation in permafrost areas. The problem is solved numerically by the finite element method. Modeling of the temperature field has been carried out for 30 years since the start of the gas pipeline operation. Calculations are carried out until the periodic temperature regime of the soil around the gas pipeline is practically established. Under the initial conditions adopted in the work, a periodic temperature regime at the top and bottom of the pipe is established in approximately 12 years, and a periodic temperature regime in depth is established in approximately 22 years.**

**Two scenarios of climate warming are considered: moderate RCP2.6 and more negative RCP8.5. It is shown that significant changes in the ground temperature regime occur in the vicinity of the pipe under both warming scenarios. Nevertheless, the calculations demonstrate the preservation of permafrost even in a negative scenario of climate warming.**

## I. Introduction

In [1], we consider an initial-boundary value problem for a nonstationary two-dimensional heat conduction equation in a bounded domain that simulates the nonstationary temperature distribution of permafrost soil in the vicinity of a main gas pipeline.

A. A. Fedotov – Bauman Moscow State Technical University (5/1 2-nd Baumanskaya St., Moscow 105005, Russia),
ORCID: https://orcid.org/0000-0003-3383-1188,
e-mail: le-tail@list.ru , fedotov_a_a@bmstu.ru.
P. V. Khrapov – Bauman Moscow State Technical University (5/1 2-nd Baumanskaya St., Moscow 105005, Russia),
ORCID: https://orcid.org/0000-0002-6269-0727,
e-mail: pvkhrapov@gmail.com, khrapov@bmstu.ru.
A.E. Dengovskaya – Bauman Moscow State Technical University (5/1 2-nd Baumanskaya St., Moscow 105005, Russia),
e-mail: alenadengovskaya@yandex.ru.

The parameters of the mathematical model are selected in accordance with experimental data on gas transportation in permafrost areas. Modeling of the temperature field is carried out for 30 years from the start of operation of the gas pipe. It is shown that the operation of the pipe with the selected parameters does not lead to the appearance of areas with a positive temperature in the permafrost soil around the pipe. The steady-state temperature regime ensures trouble-free operation of the main gas pipeline.

In this work, the methodology from [1] is used to analyze the temperature regime of the soil in the vicinity of the section of the main gas pipeline Kysyl - Syr - Yakutsk, taking into account climate warming. When determining the parameters of the mathematical model, the following were used: data obtained from a meteorological station [2]; construction standards and rules for the operation of engineering structures in permafrost conditions [3], [4], as well as the results of engineering geological surveys [5]. Two models of climate warming are considered: RCP2. 6 corresponds to a scenario in which $CO_2$ emissions into the atmosphere were maximal in 2010-2020 and will continue to decrease until 2100; RCP8.5 implies that carbon dioxide emissions will continue to increase until 2100. In 2080, RCP2.6 forecasts an increase in average January temperatures of 3.4°C and July average temperatures of 1.9°C. In the more negative RCP8.5 scenario, winter temperatures are increased by 9.1°C, and summer temperatures are increased by 5.7°C [6].

It is shown that significant changes in the ground temperature regime occur in the vicinity of the pipe under both warming scenarios. Nevertheless, the calculations demonstrate the preservation of permafrost even in a negative scenario of climate warming.

## II. Problem statement

A section of the Kysyl – Syr – Yakutsk main gas pipeline was selected for the study. The pipeline has a total length of more than 900 km. Type of gas pipeline according to the method of laying: underground. The extended section of the gas pipeline is considered. Therefore, to study the ground temperature field in the vicinity of a gas pipeline in such a section, the problem can be formulated in a two-dimensional formulation.

It is required to find the distribution of soil temperature in the vicinity of a gas pipeline pipe. The thermal state of such a medium in dimensionless form is represented as follows.

Solving $u(t, x, y)$ the equation

$$C\frac{\partial u}{\partial t}=\frac{\partial}{\partial x}\left(Fo\frac{\partial u}{\partial x}\right)+\frac{\partial}{\partial y}\left(Fo\frac{\partial u}{\partial y}\right)+s \quad (1)$$

must be found in a limited area $D$ (Figure 1),

$$D = D_1 \setminus D_2, \; D_1 = \{-L_x \leq x \leq L_x, -L_y \leq y \leq 0\},$$
$$D_2 = \{x^2 + (y+y_c)^2 = R_T^2\},$$

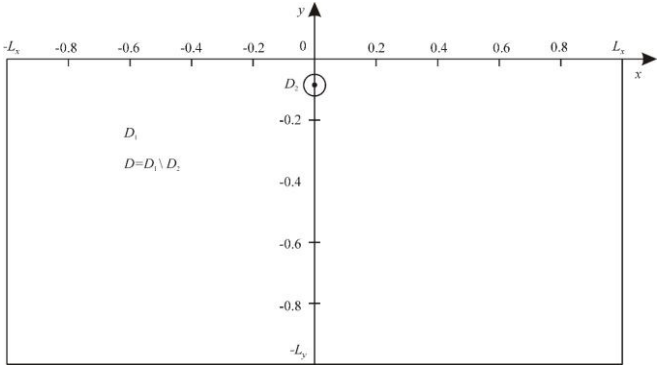

Fig. 1. Calculated area.

satisfying the initial condition

$$u(0, x, y) = \varphi(x, y). \quad (2)$$

At the upper boundary $y = 0$, convective heat exchange occurs with a medium having a temperature of $\theta(t)$

$$\frac{\partial u}{\partial y} = Bi(x, 0, t) \cdot [\theta(t) - u(t, x, 0)]. \quad (3)$$

On the lower border $y = -L_y$

$$\frac{\partial u}{\partial y} = Ki(x, -L_y, u) \quad (4)$$

The side borders of the area are heat-insulated

$$\frac{\partial u}{\partial x}(t, -L_x, y) = 0, \quad \frac{\partial u}{\partial x}(t, L_x, y) = 0 \quad (5)$$

A constant temperature is maintained on the surface of a pipe with a radius $R_T$ centered at point $O(0, y_c)$:

$$u(t, x, y) = U_p = const. \quad (6)$$

In (1)-(6): $x, y$ – Cartesian coordinates, $t$ – time, $u(t, x, y)$ – ambient temperature, $s = s(x, y, t)$ – power of internal heat sources, $C = c\rho$ – volume heat capacity, $c$ – specific heat capacity, $\rho$ – density, $Fo$ – Fourier number, $Bi$ – Bio number, $Ki$, – Kirpichev heat exchange number.

When moving to dimensionless variables, the following are taken as characteristic quantities:

$L' = \max\{L'_x, L'_y\}$ – is the maximum value of parameters $L'_x$ and $L'_y$ of the computational domain,

Here and further, the dash marks dimensional values when they need to be distinguished from the corresponding dimensionless values.

$L'_x = 20m, \; L'_y = 20m, \; L' = 20m, \; T' = 8760\,h$ - the period of change of boundary conditions at the upper boundary (1 year),

$\theta'_{max} = 40°C$ – the maximum modulo value of the temperature in the problem,

$c'_d = 0,22 \dfrac{kcal}{kg \cdot °C}$ - the specific heat capacity of dry soil, $\rho'_d = 1390 \dfrac{kg}{m^3}$ – the density of dry soil.

The volumetric heat capacity of dry soil can be taken as a characteristic value, i.e.

$$C'_d = c'_d \rho'_d = 305,8 \frac{kcal}{m^3 \cdot °C},$$

$\lambda' = 1,55 \dfrac{kcal}{m \cdot h \cdot °C}$ - coefficient of thermal conductivity.

The main designations used during the transition to a dimensionless view are:

$$x = \frac{x'}{L'}, \; y = \frac{y'}{L'}, \; t = \frac{t'}{T'}, \; u = \frac{u'}{\theta'_{max}}, \; \theta = \frac{\theta'}{\theta'_{max}},$$

$$C = \frac{C'}{C'_d} = \frac{c'\rho'}{c'_d\rho'_d} = \frac{c'_d\rho'_d}{c'_d\rho'_d} = 1,$$

$$s = \frac{s'T'}{c'_d \rho'_d \theta'_{max}}, \; s = 0, \; Fo = \frac{\lambda' T'}{c'_d \rho'_d \cdot (L')^2} \approx 0,1110,$$

$$Bi = \frac{h'L'}{\lambda'}, \; \frac{L'}{\lambda'} \approx 12.9032, \; q'_E = 0,043 \frac{kcal}{m^2 \cdot h} -$$

constant heat flow (flow from the Earth's interior [7]),

$$Ki = \frac{L' q'_E}{\lambda' \theta'_{max}} \approx 0.0139, \; U'_P = -3.5°C,$$

$$U_P = \frac{U'_P}{\theta'_{max}} = -0.0875.$$

In areas with perennially frozen ground gas is transported at a temperature of $-2°C$ in winter and no higher than $-5°C$ in summer [8], $-3.5°C$ is the average value of these temperatures.

The values of physical and thermophysical characteristics of the soil were taken in accordance with the data of [5].

### III. INITIAL AND BOUNDARY CONDITIONS

The heat transfer coefficient $h'$ is written as

$$h' = \frac{1}{\dfrac{1}{\alpha'} + R'}, \quad (7)$$

where $\alpha'$ -is the convective heat transfer coefficient, $R'$ -is the thermal resistance of the soil surface [5].

Long-term monthly averages are used to calculate the dynamics of the soil temperature regime. Data on ground temperature at various depths, air temperature, and snow cover height are taken from the archives of the Hydrometeorological Center [9] and WMO (International

Meteorological Organization) [2] for the 24643 Khatyryk – Khomo weather station (Yakutia).

The paper uses data on air and ground temperature for the period from 1970 to 2020 for the 24643 Khatyryk – Khomo weather station.

Table 1 shows the long-term average monthly values of air and ground temperature at a depth of 20, 60, 80, 120, 240 cm, snow cover height according to the 24643 Khatyryk – Khomo weather station.

Table 1. Long-term monthly average values of temperature and snow cover height.

| Month | 1 | 2 | 3 | 4 | 5 | 6 | 7 | 8 | 9 | 10 | 11 | 12 |
|---|---|---|---|---|---|---|---|---|---|---|---|---|
| Air temperature (°C) | -39,94 | -35,99 | -19,87 | -6,20 | 2,51 | 11,60 | 15,07 | 13,55 | 1,42 | -10,79 | -28,68 | -39,79 |
| Snow cover (cm) | 28,81 | 32,44 | 33,94 | 14,74 | 0,14 | 0,00 | 0,00 | 0,00 | 0,20 | 7,87 | 20,61 | 25,74 |
| 20 cm | -10,61 | -12,85 | -11,53 | -4,75 | 0,16 | 3,51 | 4,98 | 5,08 | 2,74 | -1,33 | -5,18 | -8,78 |
| 60 cm | -8,66 | -10,82 | -10,36 | -5,46 | -0,65 | 1,92 | 2,16 | 2,73 | 1,02 | -0,49 | -3,27 | -6,12 |
| 80 cm | -4,91 | -7,85 | -8,56 | -6,04 | -2,28 | -0,16 | 0,86 | 0,91 | 0,07 | -0,13 | -0,74 | -2,63 |
| 120 cm | -1,49 | -3,87 | -5,32 | -5,41 | -3,45 | -2,23 | -1,23 | -0,62 | -0,85 | -0,93 | -0,88 | -0,83 |
| 240 cm | -1,17 | -1,79 | -3,04 | -3,98 | -3,85 | -3,23 | -2,65 | -2,06 | -1,55 | -1,24 | -1,11 | -1,06 |

The thermic resistance of the snow cover is calculated by the formula [10]

$$R' = \frac{d'_s}{\lambda'_s}, \quad (8)$$

where $d'_s$ is the average monthly snow cover height (Table 1);

$\lambda'_s$ – the average monthly thermal conductivity of the snow cover, determined by the formula [10]

$$\lambda'_s = m'_d (0,18 + 0,87 \rho'_s), \quad (9)$$

where $m'_d = 0,001 \, m^2 \cdot kcal/(kg \cdot h \cdot °C)$ – is the conversion factor;

$\rho'_s$ – monthly average snow cover density, $kg/m^3$, taken from meteorological data [11].

Using data for $d'_s$, $\lambda'_s$, $\rho'_s$, [2,9-11], $\alpha'$ [5] Table 2 with physical characteristics at the upper boundary of the computational domain is compiled.

Table 2. Physical characteristics at the upper boundary of the computational domain.

| Month | 1 | 2 | 3 | 4 | 5 | 6 | 7 | 8 | 9 | 10 | 11 | 12 |
|---|---|---|---|---|---|---|---|---|---|---|---|---|
| $\theta', °C$ | -39,9 | -36,0 | -19,9 | -6,2 | 2,5 | 11,6 | 15,1 | 13,6 | 1,4 | -10,8 | -28,7 | -39,8 |
| $d'_s, cm$ | 28,81 | 32,44 | 33,94 | 14,74 | 0,14 | 0,00 | 0,00 | 0,00 | 0,20 | 7,87 | 20,61 | 25,74 |
| $\rho'_s, kg/m^3$ | 256,7 | 260 | 263,3 | 283,3 | 313,3 | 0 | 0 | 0 | 0 | 156,7 | 213,3 | 226,7 |
| $R', \frac{m^2 h °C}{kcal}$ | 2,6 | 2,7 | 2,8 | 3,0 | 1,7 | - | - | - | - | 1,3 | 2,1 | 2,3 |
| $\alpha', \frac{kcal}{m^2 h °C}$ | 13,8 | 12,3 | 12,3 | 12 | 12,1 | 12,6 | 13,4 | 14 | 14,3 | 13,4 | 13,5 | 12,7 |
| $h', \frac{kcal}{m^2 h °C}$ | 0,37 | 0,36 | 0,35 | 0,32 | 0,56 | 12,6 | 13,4 | 14 | 14,3 | 0.73 | 0,46 | 0,42 |

According to these data, approximation formulas are compiled using the least squares method [12] for the dependencies $\theta(t)$, $h(t)$. The approximation by a polynomial of degree 5 for $\theta(t)$ is constructed from the points indicated in Table 3.

Table 3. Points selected for constructing the approximation $\theta(t)$.

| $t$ | 0.0 | 0.16 | 0.496 | 0.666 | 0.915 | 1.0 |
|---|---|---|---|---|---|---|
| $\theta$ | -1.0 | -0.9 | 0.315 | 0.315 | -0.72 | -1.0 |

As a result, we got:

$$\tau_1(\bar{t}) = -1 - 2.1222\bar{t} + 20.1562\bar{t}^2 - \\ - 26.2135\bar{t}^3 - 17.8805\bar{t}^4 + 16.06\bar{t}^5. \quad (10)$$

Here $\bar{t} = [t', T']/T'$, $[t', T']$ – means the remainder of the integer division of the first argument by the second.

$$\tau_2(t) = max(-1, \tau_1) \quad (11)$$

Then the temperature is:

$$\theta(t) = min(0.315, \tau_2) \quad (12)$$

On the left in Figure 2, the graph of the temperature distribution by month is shown in black; at the points indicated in Table 3, the curve that approximates it is shown in pink. As a result, a periodic temperature-time dependence $\theta(t)$ is obtained, the graphical representation of which is shown on the right in Figure 2.

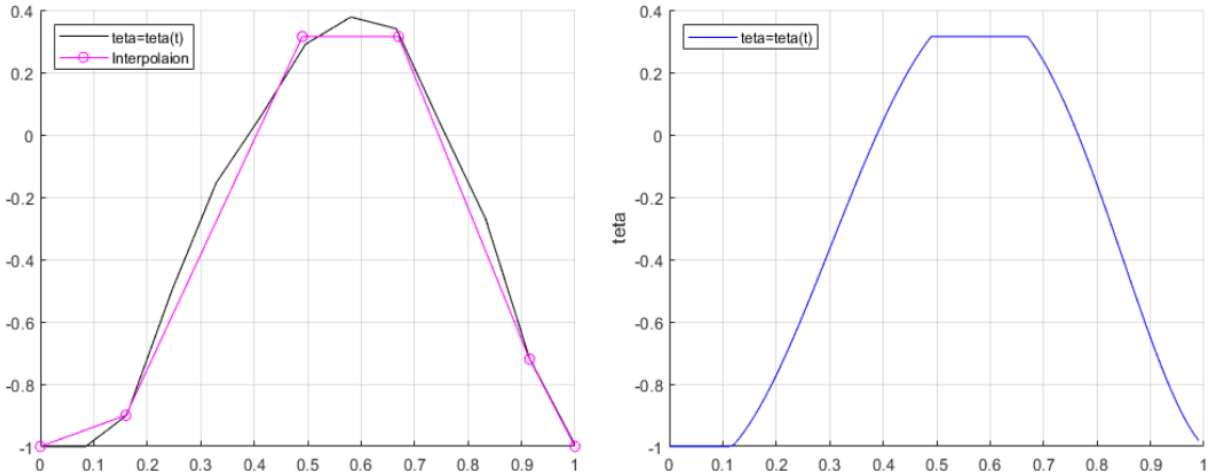

Fig. 2. Graph $\theta(t)$.

An approximation for the dependence $h(t)$ is constructed using the method described above. The approximation by a polynomial of degree 5 for $h(t)$ is constructed from the points indicated in Table 4.

Table 4. Points selected for constructing the approximation $h(t)$.

| $t$ | 0.0 | 0.41 | 0.5 | 0.752 | 0.83 | 1.0 |
|---|---|---|---|---|---|---|
| $h$ | 5.0 | 5.0 | 170 | 170 | 5.0 | 5.0 |

As a result, we got:

$$\eta_1(t) = 5 - 173\bar{t} - 13975\bar{t}^2 + \\ + 67491\bar{t}^3 - 96770\bar{t}^4 + 43428\bar{t}^5 \quad (13)$$

$$\eta_2(t) = max(5, \eta_1) \quad (14)$$

Then the temperature is:

$$h(t) = min(170, \eta_2) \quad (15)$$

On the left side of Figure 3, a graph of changes in the heat transfer coefficient by month is shown in black; at the points indicated in Table 4, a curve is plotted in pink, which approximates it. As a result, a periodic dependence of the heat transfer coefficient on time $h(t)$ is obtained, the graphical representation of which is shown on the right in Figure 3.

The initial condition (2) was assumed to be equal $-0.0875$.

The calculations were performed with the following values of the geometric characteristics of the problem: pipe radius $R'_T = 0.71$ m, then $R_T = 0.0355$. The distance from the ground surface to the pipe axis is assumed to be $(R'_T + 1.0)$ m. [13]. Coordinates of the pipe center: $(x_c = 0.0, y_c = -0.0855)$. The ratio $L_x$ to the pipe radius is: $L_x/R_T > 28$, i.e. it can be assumed that the lateral boundary conditions (5) do not affect the distribution of ground temperature in the vicinity of the pipe [1].

IV. RESULTS OF CALCULATIONS OF THE GROUND TEMPERATURE DISTRIBUTION IN THE VICINITY OF THE GAS PIPELINE

The numerical solution of equation (1) with the corresponding conditions (2)-(6) is obtained using the finite element method [14], [15].

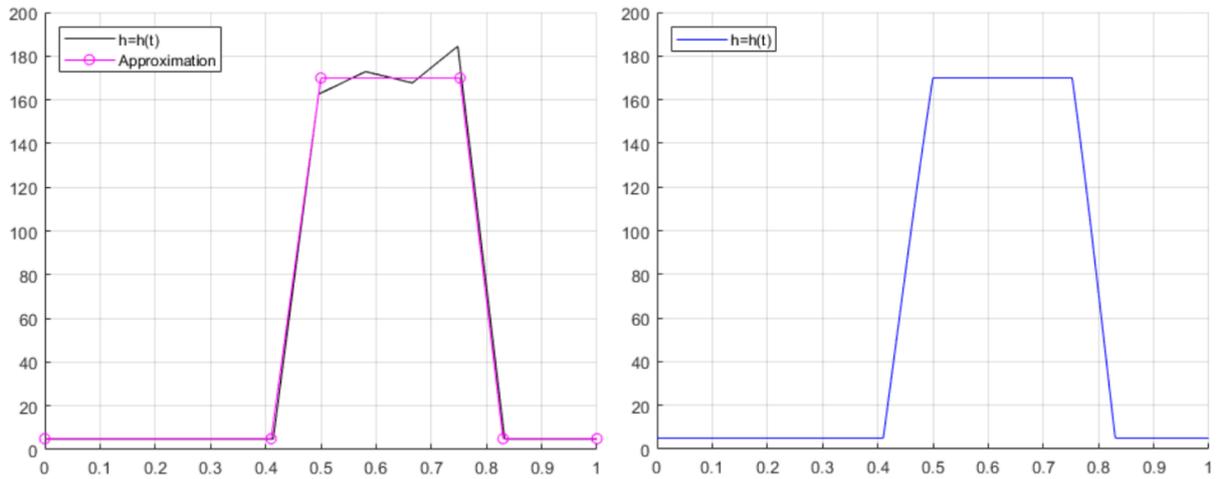

Fig. 3. Graph $h(t)$.

The calculations start from January 1 of the first year of gas pipeline operation. To analyze the temperature field in the vicinity of the pipe at certain moments of time, the temperature $u$ dependence along the $x$ axis was plotted at a distance of 10 cm from the top point of the pipe and 10 cm from the bottom point of the pipe. At a distance of 10 cm from the rightmost point of the pipe, the temperature $u$ along the $y$ axis was plotted along the depth $z = -y$.

Figure 4 shows the temperature $u = u(x)$ dependencies at a distance of 10 cm from the top point of the pipe for the middle of the 1st, 5th, 10th, 15th, 20th, and 30th years of observations.

Figure 5 shows the temperature $u = u(x)$ dependences at a distance of 10 cm from the lower point of the pipe for the same time values: for the middle of the 1st, 5th, 10th, 15th, 20th, and 30th observation years.

Figures 4 and 5 show graphs for the right half of the computational domain, since the axis $y$ is the axis of symmetry of the problem. The graphs show that the periodic temperature regime along the $x$ axis is established after about 12 years.

Figure 6 shows the temperature $u = u(z)$ dependences along the $y$ axis along the depth $z = -y$ at a distance of 10 cm from the rightmost point of the pipe for the middle of the 1st, 5th, 10th, 15th, 20th, and 30th years of observations.

It follows from the graphs in Figure 6 that the periodic distribution of temperature over depth is established after approximately 22 years.

According to Figures 4-6, it can be concluded that the ground temperature in the vicinity of the pipe remains negative throughout the entire observation period.

V. DISTRIBUTION OF PERMAFROST GROUND TEMPERATURE AROUND THE MAIN GAS PIPELINE, TAKING INTO ACCOUNT CLIMATE WARMING

Climate change modeling, including its duration and intensity, mainly uses global climate models of the general circulation of the atmosphere and ocean (AOGCM). Within the framework various international programs, including AMIP (Atmospheric Model Intercomparison Project) and CMIP (Coupled Model Intercomparison Project), compare models both with each other and with already known meteorological data.

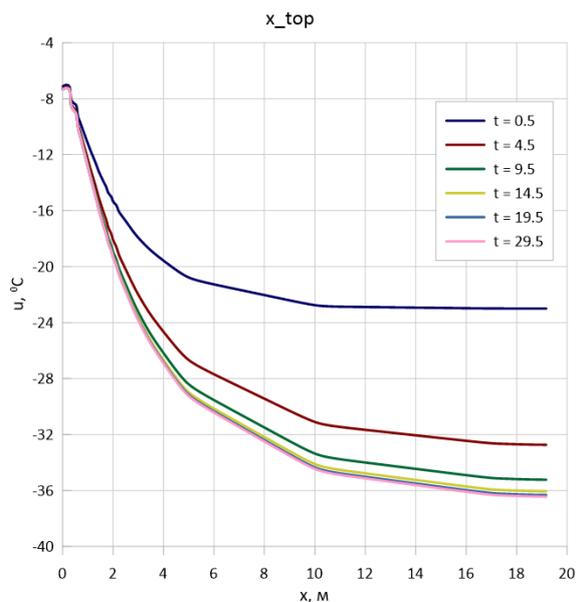

Fig. 4. Temperature distribution along the axis $x$ at a distance of 10 cm from the top point of the pipe.

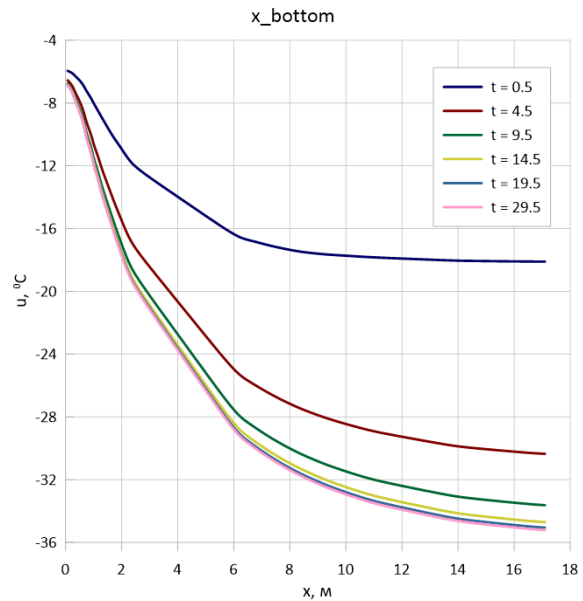

Fig. 5. Temperature distribution along the axis $x$ at a distance of 10 cm from the bottom point of the pipe.

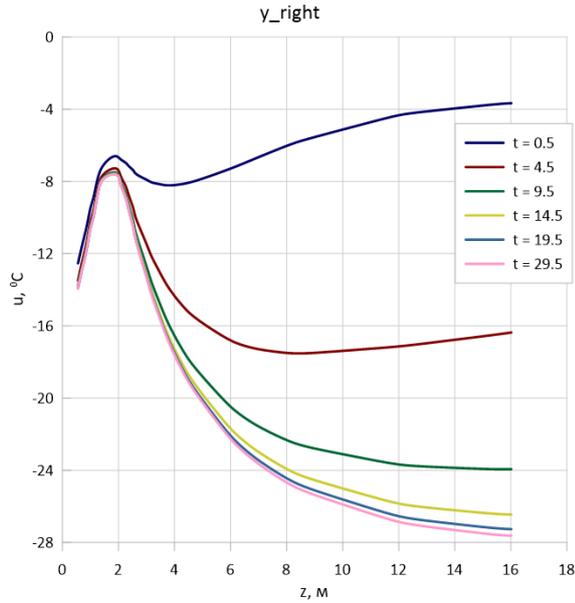

Fig. 6. Temperature distribution by depth at a distance of 10 cm from the rightmost point of the pipe.

There are ready models of climate change built within the CMIP5 project in preparation of the 5th Assessment Report of the Intergovernmental Panel on Climate Change [16]. The CMIP5 project includes scenarios for estimating greenhouse gases and other radiation substances – RCP (Representative Concentration Pathway) [6], which are the main cause of climate change.

In this paper, an ensemble of global climate models CMIP5 and two scenarios of changes in the content of greenhouse gases and aerosols in the atmosphere are used to assess climate change in the vicinity of the Kysyl – Syr – Yakutsk gas pipeline section: moderate – RCP2.6 [17] and negative – RCP8.5 [18].

According to [19] and materials of the Voeikov Geophysical Observatory [20], by the end of 2100, changes in the average air temperature described below are predicted in the Asian part of the Russian Federation, namely in Western Siberia.

The calculation results for the negative RCP 8.5 scenario are shown below.

For the RCP8.5 scenario, when moving to dimensionless variables, the same variables as in section II are taken as characteristic parameters, with the following exception:

$\theta'_{max} = 30°C$ – the maximum value of the temperature modulus in the problem.
Then
$$U_P = U'_P / \theta'_{max} = -0.117,$$
$$Ki = L' q'_E / \lambda' \theta'_{max} \approx 0.018$$

Table 5 shows the monthly average values of ambient temperature $\theta'$ for the RCP8.5 scenario. The approximation by a degree 5 polynomial for $\theta(t)$ is plotted from the points indicated in Table 6.

As a result, we got:
$$\tau_1(t) = -1 - 4.481\bar{t} + 46.2984\bar{t}^2 - \\ - 94.1501\bar{t}^3 + 76.0541\bar{t}^4 - 23.7215\bar{t}^5, \quad (16)$$
$$\tau_2(t) = max(-1, \tau_1) \quad (17)$$

Then the temperature is:
$$\theta(t) = min(0.569, \tau_2) \quad (18)$$

On the left side of Figure 7 are plots of the temperature distribution $\theta(t)$ by month at present and under the RCP8.5 scenario: the black graph corresponds to the present, the blue one to the RCP8.5 scenario. Points are selected on each of the graphs and curves are constructed according to them, in pink and light green, respectively, which approximate them. As a result, periodic dependences of temperature $\theta(t)$ on time are obtained, shown on the right side of Figure 7. In calculations for the RCP8.5 scenario, the function $h(t)$ is represented by formula (15) (Figure 3).

VI. RESULTS OF CALCULATIONS OF THE PERMAFROST GROUND TEMPERATURE DISTRIBUTION AROUND THE MAIN GAS PIPELINE, TAKING INTO ACCOUNT CLIMATE WARMING

When predicting changes in the ground temperature regime under the negative warming scenario RCP8.5, the following results were obtained.

Figure 8 shows the temperature $u = u(x)$ dependencies at a distance of 10 cm from the top point of the pipe for the middle of the 1st, 5th, 10th, 15th, 20th, and 30th years of observations.

Figure 9 shows the temperature $u = u(x)$ dependences at a distance of 10 cm from the lower point of the pipe for the same time values: for the middle of the 1st, 5th, 10th, 15th, 20th, and 30th observation years.

Table 5. Average monthly air temperature of the RCP8.5 scenario

| Month | 1 | 2 | 3 | 4 | 5 | 6 | 7 | 8 | 9 | 10 | 11 | 12 |
|---|---|---|---|---|---|---|---|---|---|---|---|---|
| $\theta', °C$ | -30.2 | -26.3 | -13.4 | 0.3 | 9.0 | 17.2 | 20.7 | 19.2 | 9.4 | 2.8 | -20.7 | -30.1 |

Table 6. Points selected for constructing the approximation $\theta(t)$.

| $t$ | 0 | 0.16 | 0.496 | 0.666 | 0.833 | 1 |
|---|---|---|---|---|---|---|
| $\theta$ | -1 | -0.87 | 0.569 | 0.569 | 0.092 | -1 |

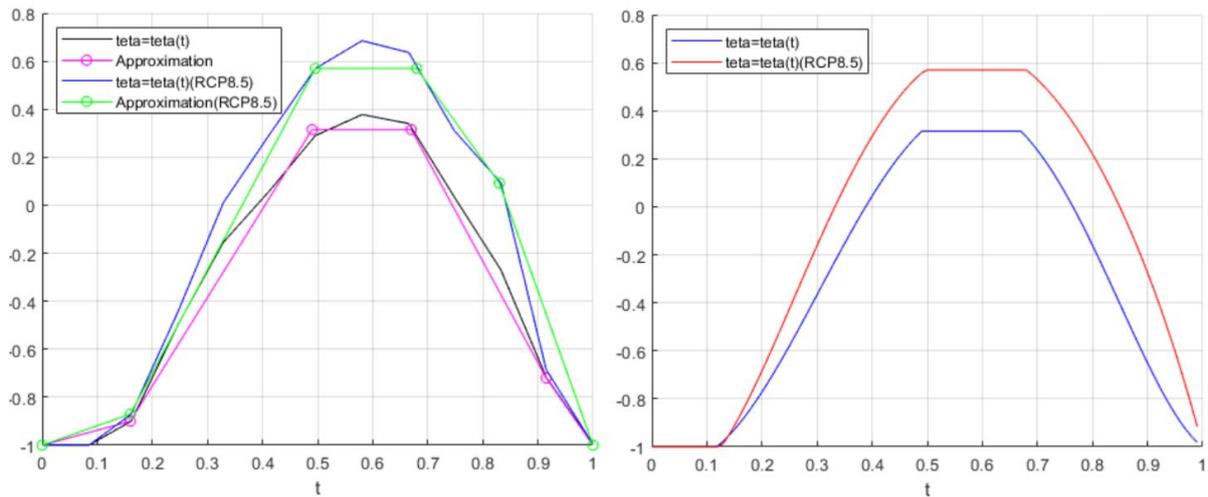

Fig. 7. Graphs $\theta(t)$ for the RCP8.5 scenario.

Figures 8 and 9, as above, show the results for the right half of the computational domain.

The graphs in Figures 8 and 9 show that the periodic temperature regime along the $x$ axis is established after approximately 12 years.

Figure 10 shows the temperature $u = u(z)$ dependences along the $y$ axis along the depth $z = -y$ at a distance of 10 cm from the rightmost point of the pipe for the middle of the 1st, 5th, 10th, 15th, 20th, and 30th years of observations.

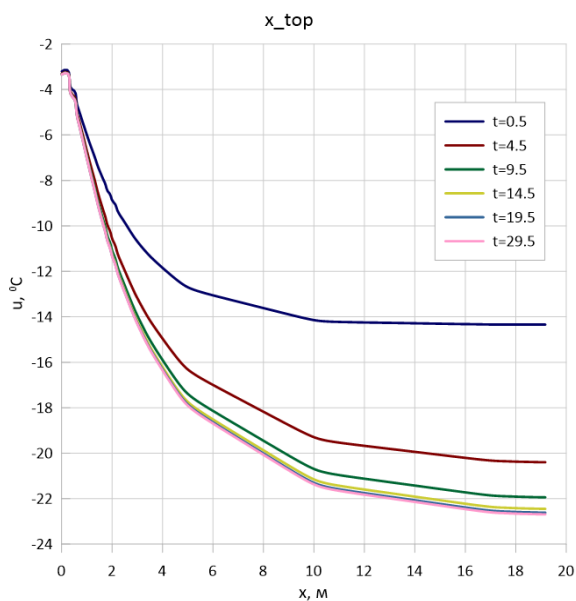

Fig. 8. Temperature distribution along the axis $x$ at a distance of 10 cm from the top point of the pipe.

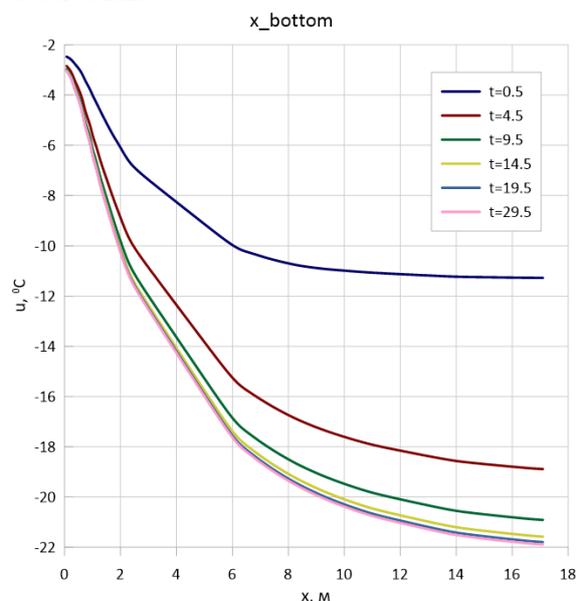

Fig. 9. Temperature distribution along the axis $x$ at a distance of 10 cm from the bottom point of the pipe.

It follows from the graphs in Figure 10 that the periodic distribution of temperature over depth is established after approximately 22 years.

Figure 11 shows the temperature field $u = u(t, x, y)$ at the end of the 30th year of observations.

According to Figures 8-11, it can be concluded that the ground temperature in the vicinity of the pipe remains negative throughout the entire observation period. Comparison with the results of calculations shown in Figure 6 shows that the established periodic temperature regime in the RCP8.5 scenario at a depth of 16-20 m shifts by about $9°C$ upwards, which already significantly increases the risks of an emergency situation in the future.

Similar curves calculated for the RCP2.6 scenario showed that the ground temperature in the vicinity of the pipe also remains negative throughout the entire observation time. In this case, from the comparison with the calculation results shown in Figure 6, the upward bias of the steady-state periodic temperature regime under the RCP2.6 warming scenario at 16 -20 m depth is more moderate at about $5°C$.

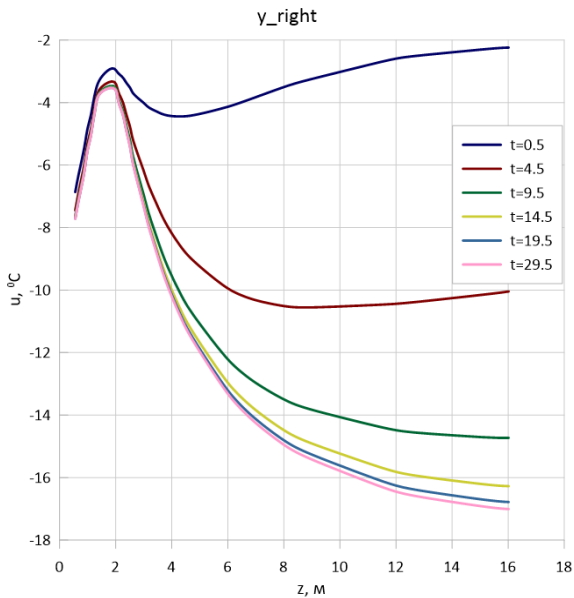 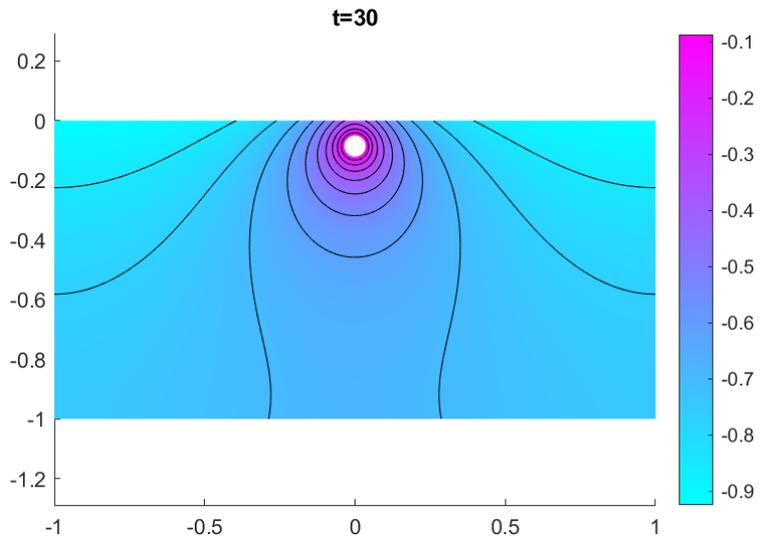

Fig. 10. Temperature distribution by depth at a distance of 10 cm from the rightmost point of the pipe.

Fig. 11. Temperature distribution at $t = 30$.

## VII. CONCLUSION

The analysis of the ground temperature field around the Kysyl-Syr-Yakutsk main gas pipeline has been carried out. The results obtained allow us to conclude that the operation of the gas pipeline with the specified parameters under the existing ambient temperature regime is safe. Thawing of soil in the vicinity of the gas pipeline does not occur. Steady state periodic temperature regimes allow us to draw conclusions about the state of the ground temperature field in the vicinity of the main gas pipeline at the present time and in the future, taking into account climate change.